\documentclass[a4paper,fleqn]{cas-dc}

\usepackage[authoryear]{natbib}

\usepackage{bm} % for bold maths
\def\tsc#1{\csdef{#1}{\textsc{\lowercase{#1}}\xspace}}
\tsc{WGM}
\tsc{QE}
\tsc{EP}
\tsc{PMS}
\tsc{BEC}
\tsc{DE}
\begin{document}
\let\WriteBookmarks\relax
\def\floatpagepagefraction{1}
\def\textpagefraction{.001}
\shorttitle{Tidal controls on Io's lithospheric thickness and topography}
\shortauthors{DC Spencer et~al.}

\title [mode = title]{Tidal controls on the lithospheric thickness and topography of Io from magmatic segregation and volcanism modelling}

\author[1]{DC Spencer}[orcid=0000-0003-1814-964X]
\cormark[1]
\ead{dan.spencer@earth.ox.ac.uk}

\address[1]{Department of Earth Sciences, University of Oxford, Oxford, UK}

\author[1]{RF Katz}[type=editor,
                        orcid=0000-0001-8746-5430]
\ead{richard.katz@earth.ox.ac.uk}

\author[2]{IJ Hewitt}[orcid=0000-0003-1814-964X]
\ead{hewitt@maths.ox.ac.uk}

\address[2]{Mathematical Institute, University of Oxford, Oxford, UK}

\cortext[cor1]{Corresponding author}

\begin{abstract}
Tidal heating is expected to impart significant, non-spherically-symmetric structure to Jupiter's volcanic moon Io. A signature of spatially variable tidal heating is generally sought in observations of surface heat fluxes or volcanic activity, an exploration complicated by the transient nature of volcanic events. The thickness of the lithosphere is expected to change over much longer timescales, and so may provide a robust link between surface observations and the tidal heating distribution. To predict long-wavelength lithospheric thickness variations, we couple three-dimensional tidal heating calculations to a suite of one-dimensional models of magmatic segregation and volcanic eruption. We find that the lithospheric thickness could either be correlated with the radially integrated heating rate, or weakly anti-correlated. Lithospheric thickness is correlated with radially integrated heating rate if magmatic intrusions form at a constant rate in the lithosphere, but is weakly anti-correlated if intrusions form at a rate proportional to the flux through volcanic conduits. Utilising a simple isostasy model we show how variations in lithospheric thickness can predict long-wavelength topography. The relationship between lithospheric thickness and topography depends on the difference in chemical density between the lithosphere and mantle. Assuming that this difference is small, we find that long-wavelength topography anti-correlates with lithospheric thickness. These results will allow future observations to critically evaluate models for Io's lithospheric structure, and enable their use in constraining the distribution of tidal heating.
\end{abstract}

%\begin{graphicalabstract}
%\includegraphics{figs/grabs.pdf}
%\end{graphicalabstract}

\begin{highlights}
\item Io is modelled with a spatially-variable tidal heating model coupled to a magma-segregation and volcanism model.
\item We predict long-wavelength lithospheric thickness variations that arise from spatially variable tidal heating.
\item Lithospheric thickness could either be correlated with surface heat flux (if intrusions form at a constant rate), or be weakly anti-correlated (if intrusions form at a rate proportional to magma flux)
\item An isostasy model predicts long-wavelength topography, showing that if the lithosphere and upper mantle have similar chemical densities, topography is predicted to anti-correlate with lithospheric thickness.
\end{highlights}

\begin{keywords}
Tidal heating \sep Volcanism \sep Magmatism \sep Heat-piping \sep Geodynamics
\end{keywords}

\maketitle

\section{Introduction}

Io, the most volcanic body in the solar system, operates in a different tectonic regime to the terrestrial planets. The eruption and burial of lava, combined with the low surface temperature leads to the growth of a thick and cold lithosphere in spite of the high surface heat flux. This high heat flux is primarily exported from the interior by magmatic segregation in the mantle \citep{moore_thermal_2001}, and through volcanic `heat-pipes' in the lithosphere \citep{oreilly_magma_1981}. Heat is supplied to the interior by tidal dissipation --- a process of great importance in the Solar System \citep{de_kleer_tidal_2019} --- and the distribution of input tidal heating is expected to control the surface heat flux distribution \citep{ross_internal_1990,tackley_convection_2001,kirchoff_global_2011,beuthe_spatial_2013,rathbun_global_2018,steinke_tidally_2020}. However, the implications of tidal heating for interior structure, magma dynamics, tectonics and topography are not well known.

The spatial distribution of tidal heating within a body is a longstanding and still largely unresolved problem in planetary science \citep{segatz_tidal_1988,roberts_tidal_2008,beuthe_spatial_2013,bierson_test_2016,renaud_increased_2018}. The end-members generally considered for Io are that of lower-mantle heating or asthenosphere heating \citep{segatz_tidal_1988,ross_internal_1990,tackley_three-dimensional_2001,hamilton_spatial_2013}, though magma-ocean dissipation has also been proposed \citep{tyler_tidal_2015,hay_powering_2020}. Lower-mantle dissipation predicts high polar heat fluxes, whereas asthenospheric dissipation predicts high equatorial heat fluxes. A number of works have sought to identify tidal dissipation patterns from surface heat fluxes \citep{veeder_io:_2012}, volcanic activity \citep{rathbun_global_2018}, and volcano distributions \citep{ross_internal_1990,kirchoff_global_2011,hamilton_spatial_2013}. The primary hindrance to these works is the poor polar coverage of observations, so whilst a number of these works favour an asthenosphere heating model (e.g., \cite{ross_internal_1990,kirchoff_global_2011}), the general consensus is that more polar observations are needed to fully address this question \citep{rathbun_global_2018,de_kleer_tidal_2019}. Furthermore, long-timescale, averaged heat fluxes are difficult to estimate, and it is unclear to what extent short-timescale observations of volcanic activity reflect the global dissipation structure. Tectonic features, which vary on much longer timescales, may provide a more robust link between surface observations and the distribution of tidal heating. 

An important tectonic feature that is expected to relate to the surface heat flux, and thus the tidal heating distribution, is the long-wavelength lithospheric thickness \citep{ross_internal_1990,steinke_tidally_2020}. Recent studies have proposed two hypotheses for the primary controls on the thickness of the lithosphere, which we define as the upper-most, fully solid layer of Io. \cite{steinke_tidally_2020} proposed a stagnant lid convection model where a portion of mantle heat transport occurs by convection, and that this convective heat is transported through the lithosphere by conduction. This predicts that the lithosphere is thinnest where heat flux is highest. Alternatively, \cite{spencer_magmatic_2020} proposed that the eruption and burial of lava results in the growth of a cold lithosphere, with a steady-state thickness that is controlled by the balance of downward advection and heat delivered by magmatic intrusions. Conduction plays a minor role in this model because the rate of burial is so large. In such a system, the lithospheric thickness is primarily controlled by the rate of melt production and the rate of intrusive heating. It should be noted that both of these previous studies referred to the surface layer as the `crust'; here we use `lithosphere' instead because the crust is usually considered to be a petrologically distinct layer, a distinction that becomes important in the isostatic calculations below. If the thickness of the lithosphere can be related to topography and heat flow, then long-wavelength variations in lithospheric thickness could be used to infer the tidal heating distribution. Constraints on the lithospheric thickness could also be combined with future spacecraft measurements of Io's libration amplitude to investigate its interior structure \citep{vanhoolst_librations_2020}.

Each of the models described above propose different controls on the lithospheric thickness and so may be expected to predict different relationships between the tidal heating distribution and thickness. \cite{steinke_tidally_2020} used radially integrated tidal heating to predict the effect of spatially variable heat input on lithospheric thickness, finding that the thickness anti-correlates with surface heat flux. In the present work we extend the model of \cite{spencer_magmatic_2020} to consider the effect of variable tidal heating on the eruption and intrusion model for lithospheric thickness, such that comparisons can be made between the models of \cite{spencer_magmatic_2020} and \cite{steinke_tidally_2020}.

We generalise the simplified, steady-state model of \cite{spencer_magmatic_2020} to allow variable tidal heating. Io is divided into a set of laterally contiguous, one-dimensional columns that are coupled to a viscoelastic tidal heating model. The tidal heating model calculates a three-dimensional heating rate from a spherically symmetric rheological structure. This leads to a recognised limitation of these type of tidal heating models; the three-dimensional heating rate that they produce generates a non-spherically symmetric structure that cannot be used to recalculate the heating distribution without averaging over spherical shells \citep{roberts_tidal_2008,bierson_test_2016}. Thus, models coupling such tidal heating calculations to dynamics cannot be fully self-consistent. We use this coupled, pseudo-three-dimensional model to investigate the links between tidal heating, lithospheric thicknesses, and long-timescale eruption rates/heat fluxes. Our results show that the relationship between lithospheric thicknesses and heat flux depends on how magmatic intrusions form within the lithosphere. If the rate of formation of permanent magmatic intrusions is independent of the (non-zero) magma flux through volcanic conduits, as may be expected if the volcanic system exploits pre-existing fractures, we predict the lithosphere to be thickest where radially integrated heating rate (and thus eruption rate and heat flux) is highest. If, however, magmatic intrusions form at a rate proportional to the magma flux in volcanic conduits, as may be expected if volcanic conduits form due to basal magma pressure that generates new pathways for magma to propagate into the lithosphere, the lithospheric thickness should be weakly anti-correlated with radially integrated heating rate. 

Having predicted the lithospheric thickness and its global variation, we then use a simple isostasy calculation to convert it to topography. This calculation assumes that the density difference between the lithosphere and mantle, which depends on both temperature and composition, is known. If the lithosphere is assumed to be compositionally similar to the mantle, thermal effects control density variations, and are such as to predict that topography anticorrelates with lithospheric thickness. If, on the other hand, the lithosphere is assumed to be compositionally distinct and of a lower chemical density, topography is predicted to correlate with thickness. These isostasy results are an independent extension to the lithospheric thickness calculation; the lithospheric thickness calculations do not require assumptions about the compositions or densities of the lithosphere and mantle. The isostasy model relates a feature that generally has to be indirectly inferred (lithospheric thickness), to an observation that is more readily obtained (topography). Improved observations of surface heat fluxes and their relationships to lithospheric thickness and topography will test different models for the controls on Io's lithospheric thickness \citep{steinke_tidally_2020,spencer_magmatic_2020}. With a means of critically evaluating these models, the structure of tidal heating can feasibly be constrained by future estimates of lithospheric thickness.

\section{Methodology}
Our model consists of two main parts: a theory for magmatic segregation and volcanism, and another for tidal dissipation. We append a separate isostasy calculation. The one-dimensional, magmatic segregation and volcanism model is a generalisation of the asymptotic approximation in \cite{spencer_magmatic_2020}. In it, melting is driven by the calculated tidal dissipation, which most closely follows the approach of \cite{beuthe_spatial_2013}, utilising a Maxwell formulation of viscoelasticity. Rheological parameters required by the tidal calculation are predicted by the segregation and volcanism model, completing the coupling of the two systems. The isostasy calculations utilise the equal-pressure formulation of \cite{hemingway_isostatic_2017}.

The dynamics are described by the magmatic segregation and volcanism model. \cite{spencer_magmatic_2020} derive a system where tidal heating causes the formation of magma in the mantle that rises buoyantly toward the solid lithosphere (termed crust in that work). High magma overpressure just below the base of the lithosphere facilitates a transfer of magma from the pore space into a lithospheric magmatic plumbing system, which can be thought of as a system of dikes. Magma rising in this plumbing system can freeze into the cold, surrounding lithosphere, forming permanent magmatic intrusions, delivering both mass and energy to the surroundings. The rest of the magma in the plumbing system rises to the surface and erupts, imparting a  compensating downward flux of the (now cold) erupted products. \cite{spencer_magmatic_2020} found that the delivery of heat from the freezing of magmatic intrusions is required to raise the temperature of cold, downwelling lithosphere such that a lithospheric thickness within observational constraints can be maintained. This concept of the emplacement of permanent magmatic intrusions is an important one in the present work and is discussed below.

The dynamic model is coupled to tidal dissipation to yield a consistent, three-dimensional structure. This structure is averaged over spherical shells to give spherical symmetry, and used to calculate a three-dimensional tidal heating rate. The heating rate distribution (which importantly is not spherically symmetric) is applied to a suite of column models, producing a new three-dimensional structure. This processes is iterated until the heating-distribution converges, yielding the three-dimensional structures presented in this work. We utilise a Maxwell viscoelastic law despite the well-documented inability of such a rheological law to produce observed dissipation rates at realistic mantle viscosities \citep{bierson_test_2016,renaud_increased_2018}. We also neglect all lateral flow, justified by the long wavelength of the tidal forcing; the one-dimensional columns are considered isolated. This is a significant simplification that we discuss below, and we note that future work should aim to analyse the propensity for lateral flow. We also inherit some of the assumption of \cite{spencer_magmatic_2020}, namely that we ignore the chemical composition and as a consequence neglect the possibility of compositional convection in the mantle. Parameter values are given in table \ref{table:parameters1}.

\subsection{Magmatic segregation and volcanism}
\label{section:seg-volc}
The model of \cite{spencer_magmatic_2020} is based on conservation equations for mass, momentum, and energy in a compacting two-phase medium, together with conservation of mass in a magmatic plumbing system that transports magma through the solid lithosphere. Here, we make use of the simplified model described in appendix B of that paper. In the mantle, which is at the melting temperature $T_{m}$, tidal heating produces melt, and mass conservation of the melt phase reads
\begin{equation}
\frac{1}{r^{2}}\frac{\partial}{\partial r}\left( r^{2}q\right) = \frac{\psi}{\rho_{m} L}
\end{equation}
where $q = K_{0}\phi^{n}\Delta \rho g/\eta_{l}$ is the Darcy segregation flux related to the porosity $\phi$, where $\psi$ is the local volumetric heating rate (see section 2.2) and $L$ is the latent heat. Here $K_{0}$ is a permeability constant, $n$ is the permeability exponent, $\Delta \rho$ is the density difference between the solid and liquid, and $\eta_{l}$ is the magma viscosity (numerical values for these and other parameters are listen in table \ref{table:parameters1}). The magma flux is therefore
\begin{equation}
q = \frac{1}{\rho_{m} L}\frac{1}{r^{2}}\int_{r_{m}}^{r}\psi r^{2}~\textrm{d}r,
\label{eq:q_mantle}
\end{equation}
where $\rho_{m}$ is the density of the mantle, and $r_{m}$ is the base of the mantle. At the base of the lithosphere this flux is transferred to the plumbing system, in which the flux is denoted $q_{p}$. Conservation of mass and energy in the lithosphere are described by
\begin{subequations}
\begin{equation}
\frac{1}{r^{2}}\frac{\partial}{\partial r}\left(r^{2}(u+q_{p})\right)=0,
\label{eq:1Dmass_cont}
\end{equation}
\begin{equation}
\frac{1}{r^{2}}\frac{\partial}{\partial r}\left(r^{2}q_{p}\right) = -M,
\label{eq:1Dmass_plum}
\end{equation}
\label{eq:1Dmass}
\end{subequations}
and
\begin{equation}
\frac{1}{r^{2}}\frac{\partial}{\partial r}\left(r^{2}uT\right) = \frac{1}{r^{2}}\frac{\partial}{\partial r}\left(r^{2}\kappa\frac{\partial T}{\partial r}\right) + \frac{\psi}{\rho_{m} C} + M\left(T_{m}+\frac{L}{C}\right),
\label{eq:1Dheat}
\end{equation}
where $u$ is the solid velocity, $T$ is the temperature, $M$ is the emplacement rate (the rate at which magmatic intrusions remove material from the plumbing system), and $C$ is the specific heat capacity. The final term in equation \eqref{eq:1Dheat} represents the heating that emplacement provides to the downwelling lithosphere. The solution of equations \eqref{eq:1Dmass}--\eqref{eq:1Dheat} together determines the temperature profile in the lithosphere as well as the lithospheric thickness (see \cite{spencer_magmatic_2020} for details).

In \cite{spencer_magmatic_2020}, we assumed a temperature-dependent parametrisation of the emplacement rate $M$, but this assumption is problematic in the present case. With a coupled calculation of the tidal heating rate $\psi$, there is very little tidal heating of the cold lithosphere, and hence more heat is required from emplacement to limit the growth of the lithosphere. Using the temperature-dependent form for emplacement, we find that there is too little heating of the lithosphere to avoid it becoming unreasonably thick. We therefore consider two alternative parametrisations of the emplacement rate. First, that magmatic emplacement is at a constant rate, and second, that magmatic emplacement is a function of the amount of material in the plumbing system. To allow for both possibilities we take the emplacement rate to be
\begin{equation}
    M = \lambda_{c} + \lambda_{q}q_{p},
    \label{eq:M}
\end{equation}
and explore cases where only one of $\lambda_{c}$ or $\lambda_{q}$ is non-zero at a time. Taking a constant emplacement rate ($\lambda_{c}\neq 0$ and $\lambda_{q}=0$) can be interpreted as modelling a system of dikes where the number of dikes is fixed but the flux through them varies. If emplacement is a function of contact area with the host rock, such a system could result in emplacement rate being independent of the magma flux. This is similar to, but more simple than the temperature dependence taken in \cite{spencer_magmatic_2020}. Taking emplacement to be proportional to the amount of melt in the plumbing system ($\lambda_{c}= 0$ and $\lambda_{q}\neq 0$) can also be interpreted as a system of dikes, but where the dikes have equal fluxes and the number of dikes varies. As the flux (and thus the number of dikes) increases, the contact area with the host rock also increases, and so the total emplacement rate increases. In summary, we consider cases where emplacement is positively related to, or independent of the magma flux. We do not consider the possibility of a negative relationship between emplacement rate and magma flux as we cannot conceive of a realistic physical system that this would represent.

Finally, it is important to note that $\lambda_{c}$ and $\lambda_{q}$ parametrise long-timescale averages of a range of complex processes. As such, we do not attempt to closely interpret the numerical values of these parameters; we use values that give rise to a globally-averaged lithospheric thickness that is comparable to that inferred from observations. We focus on the broad behaviour of the model in response to these parameters. We also note from equation \eqref{eq:M} that $\lambda_{c}$ has the same units as $M$ (s$^{-1}$), whereas because $q_{p}$ multiplies a flux, it has units m$^{-1}$.

\subsection{Tidal heating}
For the calculation of tidal heating we most closely follow the methodology of \cite{beuthe_spatial_2013}. Volumetric tidal dissipation averaged over an orbit is given by \citep{tobie_tidal_2005}
\begin{equation}
    \psi(r,\theta,\varphi) = \frac{\omega_{f}}{2}\left[ \textrm{Im}(\tilde{\sigma}_{ij})\textrm{Re}(\tilde{\epsilon}_{ij}) - \textrm{Re}(\tilde{\sigma}_{ij})\textrm{Im}(\tilde{\epsilon}_{ij})\right],
    \label{eq:psi_eq}
\end{equation}
where $\omega_{f} = 4.11\times 10^{-5}~$s$^{-1}$ is the orbital frequency, $\tilde{\sigma}_{ij}$ and $\tilde{\epsilon}_{ij}$ are the components of the complex stress and strain tensors, and summation over components $i$ and $j$ is implied. We calculate the complex stress and strain tensors using the propagator matrix approach detailed in \cite{sabadini_global_2004}, \cite{beuthe_spatial_2013}, and explained in appendix A of \cite{roberts_tidal_2008}. This calculation starts with the formulation of momentum conservation and Poisson equations for a body subjected to gravitational and rotational potentials. These equations are then expanded in spherical harmonics. This results in a set of six ODEs for the radially-varying spherical harmonic coefficients, which are solved in each layer of a spherically symmetric body. Together with a rheological law these coefficients yield the complex stress and strain tensors.

The tidal potential that forces the system arises from consideration of a synchronous eccentric orbit, to first order in eccentricity. It is given by \citep{kaula_tidal_1964,tobie_tidal_2005}
\begin{equation}
\begin{split}
\Omega = r^{2}\omega_{f}^{2}e &\bigg[-\frac{3}{2}P_{2}^{0}(\textrm{cos}~\theta)\textrm{cos}(\omega_{f} t) + \frac{1}{4}P_{2}^{2}(\textrm{cos}~\theta) \\
&\left[ 3\textrm{cos}(\omega_{f} t)\textrm{cos}(2\varphi)+4\textrm{sin}(\omega_{f} t)\textrm{sin}(2\varphi)\right]\bigg],
\end{split}
\end{equation}
where $e=4.1\times 10^{-3}$ is the orbital eccentricity, $\theta$ and $\varphi$ are the colatitude and longitude (the latter being zero at the sub-Jovian point), $t$ is the time, and $P_{2}^{0}$ and $P_{2}^{2}$ are associated Legrendre polynomials.

To couple the tidal heating model to the dynamical model we follow the approach of \cite{bierson_test_2016}. We take the shear viscosity to be a function of temperature and porosity through the relationship \citep{katz_porosity-driven_2010,kelemen_review_1997}
\begin{equation}
    \eta = \eta_{0}~\textrm{exp}\left[ \frac{E_{A}}{R_{g}}\left(\frac{1}{T}-\frac{1}{T_{0}}\right) - \Lambda \phi \right],
    \label{eq:diss_eta}
\end{equation}
where $E_{A} = 3\times 10^{5}~$J/mol is the activation energy, $R_{g}$ is the gas constant, $\eta_{0}$ is a reference viscosity at the reference temperature $T_{0}$ (taken to be the melting point), and $\Lambda = 27$ is a positive constant. Temperature $T$ and porosity $\phi$ are extracted from the model in section \ref{section:seg-volc} and averaged over spherical shells, so $\eta$ depends only on radius $r$. The value of $\eta_{0}$ used is chosen so that the total global rate of tidal dissipation approximately matches the observed dissipation rate of $\sim 10^{14}~$W \citep{lainey_strong_2009}. It is well documented that a Maxwell viscoelastic constitutive law requires a very low viscosity to produce the amount of tidal heating observed in Io \citep{segatz_tidal_1988,tackley_convection_2001,bierson_test_2016,steinke_tidally_2020}. We assume that this is a failure in the present understanding of the rheology that affects dissipative processes \citep{bierson_test_2016,renaud_increased_2018}, rather than a reasonable assesment of Io's long-timescale mantle viscosity. \cite{bierson_test_2016} also take a porosity dependence of the elastic shear modulus, but we neglect this small effect in line with our simplified approach. We refer to the first coupled model, using \eqref{eq:diss_eta}, as the `mantle heating' model.

Numerous previous works have considered the possibility that tidal dissipation is concentrated within a lower-viscosity asthenosphere (e.g., \cite{segatz_tidal_1988,tackley_convection_2001,hamilton_spatial_2013,davies_map_2015}). Such a dissipative layer does not arise in the above formulation, even when a large decompacting boundary layer is included in the dynamic model \citep{spencer_magmatic_2020}, because the porosity dependence in a Maxwell viscoelastic model is too weak. In order to investigate the lithospheric thickness and long-wavelength topography implications of such a dissipation structure, we calculate an alternative `asthenospheric heating' model, where the shear viscosity in a $300~$km layer beneath the lithosphere is set to be a factor of 1000 lower than the rest of the mantle. In the asthenospheric heating model we do not include the temperature and porosity dependence of shear viscosity; in this case the heating model is decoupled from the dynamical model. We do, however, set the shear viscosity in the cold lithosphere is be effectively infinite so no dissipation occurs there, consistent with the calculated dissipation structure in the coupled mantle heating model.

The tidal heating code has been benchmarked against the radial functions in figure 2 of \cite{tobie_tidal_2005}, against the TiRADE software used in \cite{roberts_tidal_2008}, and by reproducing figures 8 and 10 of \cite{segatz_tidal_1988}.

\subsection{Isostasy calculations}
For our isostasy calculations we follow \cite{hemingway_isostatic_2017} in using an equal-pressure formulation of isostasy in spherical coordinates. This assumes that compensated columns have equal pressures at their bases (the compensation radius, $r_{ccd}$). Equal-pressure isostasy assumes that we have \citep{hemingway_isostatic_2017}
\begin{equation}
    P = \int_{r_{ccd}}^{R} \rho g~\textrm{d}r,
    \label{eq:equalP_iso}
\end{equation}
where $P$ is a constant (independent of latitude and longitude), $\rho$ is the density profile, and $R$ is local planetary radius. We take gravity $g$ to be uniform for simplicity, a reasonable assumption given the likely heavy core. We assume that density in the lithosphere is a function of temperature only
\begin{equation}
    \rho = \rho_{l}[1-\alpha(T - T_{m})],
    \label{eq:density}
\end{equation}
where $\alpha=3\times 10^{-5}~$K$^{-1}$ is the coefficient of thermal expansion, and $\rho_{l}$ is the reference lithosphere density at the melting temperature $T_{m}$. It is at this point that the distinction between the crust and lithosphere becomes important for Io. In terrestrial systems, the base of the crust represents a petrological boundary between the low-density crust and the high-density mantle. \cite{spencer_compositional_2020} proposed that efficient recycling of erupted material back into the partially molten mantle removes any significant compositional variation across this boundary. In such a view there is no petrologically distinct crust, and so there is no compositionally derived density change between the lithosphere and upper mantle. Consistent with \cite{spencer_compositional_2020}, we therefore take $\rho_{l} = \rho_{m}$ for our initial topography calculations. It is plausible, however, that certain chemical species are melted and mobilised at lower temperatures, potentially resulting in a density stratification, even if the bulk of the lithospheric material is efficiently recycled into the mantle. We investigate the effect of an upper-most layer with a density $\rho_{l} \neq \rho_{m}$ on topography in Appendix \ref{Appendix}.

The integral in equation \eqref{eq:equalP_iso} can be split at the base of the lithosphere, which has a thickness $l$ to write
\begin{equation}
P = \rho_{m}g(R-l-r_{ccd}) + \int_{0}^{l}\rho g ~\textrm{d}z, 
\end{equation}
where $z=R-l$ is the distance downward from the surface. Both $l$ and $\rho$ (in terms of temperature $T$) are known from the magmatic segregation and volcanism model, so this expression can be re-arranged to determine the variable radius $R$ relative to its spatial average $\overline{R}$. Since $P$ and $r_{ccd}$ are constant, we obtain the topography $h=R-\overline{R}$ as
\begin{equation}
h = l - \int_{0}^{l}\frac{\rho}{\rho_{m}}~\textrm{d}z + \textrm{constant},
\end{equation}
where the constant is chosen to make the spatial average of $h$ zero.

\begin{table*}[h!]
\caption{Model parameters}
\centering
\resizebox*{!}{\dimexpr\textheight-3\baselineskip\relax}{
\begin{tabular}{l l l l}
\hline
Quantity & Symbol & Preferred Value & Units \\
\hline
\bf{Dynamics model} & & & \\
Radial position & $ r $ & & m \\
Radius & $R$ & $1820$ & km \\
Core radius$^{1}$ & $r_{m}$ & $700$ & km \\
Solid velocity & $ u $ & & m/s \\
Segregation flux & $ q $ & & m/s \\
Volcanic plumbing flux & $ q_{p} $ & & m/s \\
Porosity & $ \phi $ & & \\
Permeability constant$^{2}$ & $ K_{0} $ & $10^{-7}$&m$^{2}$ \\
Permeability exponent$^{2}$ & $n$ & 3 & \\
Reference mantle density & $\rho_{m} $ & $3000$ & kg/m$^{3}$ \\
Solid--liquid density difference & $\Delta \rho $ & $500$ & kg/m$^{3}$ \\
Gravitational acceleration & $g$ & $1.5$ & m/s$^{2}$ \\
Liquid viscosity & $\eta_{l}$ & $1$& Pas \\ 
Emplacement rate & $M$ & & s$^{-1}$ \\
Emplacement constant$^{*}$ & $\lambda_{c}$ & 1.66 & Myr$^{-1}$ \\
Emplacement constant$^{*}$ & $\lambda_{q}$ & 0.05 & km$^{-1}$ \\
Temperature & $T$ & & K \\
Melting temperature & $T_{m}$ & $1500$ & K \\
Surface temperature & $T_{s}$ & $150$ & K \\
Latent heat & $L$ & $4\times 10^{5}$ & J/kg \\
Specific heat capacity & $C$ & $1200$ & J/kg/K \\
Thermal diffusivity & $\kappa$ & $10^{6}$ & m$^{2}$/s \\
\hline
\bf{Tidal heating model} & & & \\
Colatitude & $ \theta $ & & rad \\ 
Longitude & $ \varphi $ & & rad \\
Orbital frequency & $ \omega_{f} $ & $4.11\times 10^{-5}$ & s$^{-1}$ \\
Orbital eccentricity & $ e $ & $4.1\times 10^{-3}$ & \\ 
Time & $ t $ & & s \\
Complex stress tensor & $ \tilde{\bm{\sigma}} $ & & Pa \\
Complex strain tensor & $ \tilde{\bm{\epsilon}} $ & & \\ 
Tidal potential & $\Omega$ & & m$^{2}$s$^{-2}$ \\
Associated Legendre polynomial & $P_{l}^{m}(x)$ & & \\
Shear viscosity & $ \eta $ & & Pas \\ 
Reference shear viscosity$^{**}$ & $ \eta_{0} $ & & Pas \\ 
Activation energy & $ E_{A} $ & $3\times 10^{5}$ & J/mol  \\ 
Reference temperature & $ T_{0} $ & & K \\ 
Porosity constant & $ \Lambda $ & 27 & \\
Tidal heating rate & $\psi$ & & W/m$^{-3}$ \\
\hline
\bf{Isostasy model} & & & \\
Pressure & $P$ & & Pa \\
Depth & $z$ & & km \\
Compensation depth & $r_{ccd}$ & & km \\
Reference lithosphere density & $\rho_{l} $ & $\rho_{m}$ & kg/m$^{3}$ \\
Lithosphere density & $\rho $ & & kg/m$^{3}$ \\
Thermal expansivity & $\alpha$ & $3\times 10^{-5}$ & K$^{-1}$ \\
Lithospheric thickness & $l$ & & km \\
Topography & $h$ & & km \\
\hline
\multicolumn{4}{l}{$^{1}$\cite{bierson_test_2016}, $^{2}$\cite{katz_magma_2008}, $^{3}$\cite{lainey_strong_2009}} \\
\multicolumn{4}{l}{$^{*}$ Chosen to give an average lithospheric thickness of $\sim 35~$km} \\
\multicolumn{4}{l}{$^{**}$ Chosen to give a total heating rate of $10^{14}~$W} \\
\end{tabular}
}
\label{table:parameters1}
\end{table*}

\section{Results and discussion}
\begin{figure*}[ht!]
    \centering
    \includegraphics[width=\linewidth]{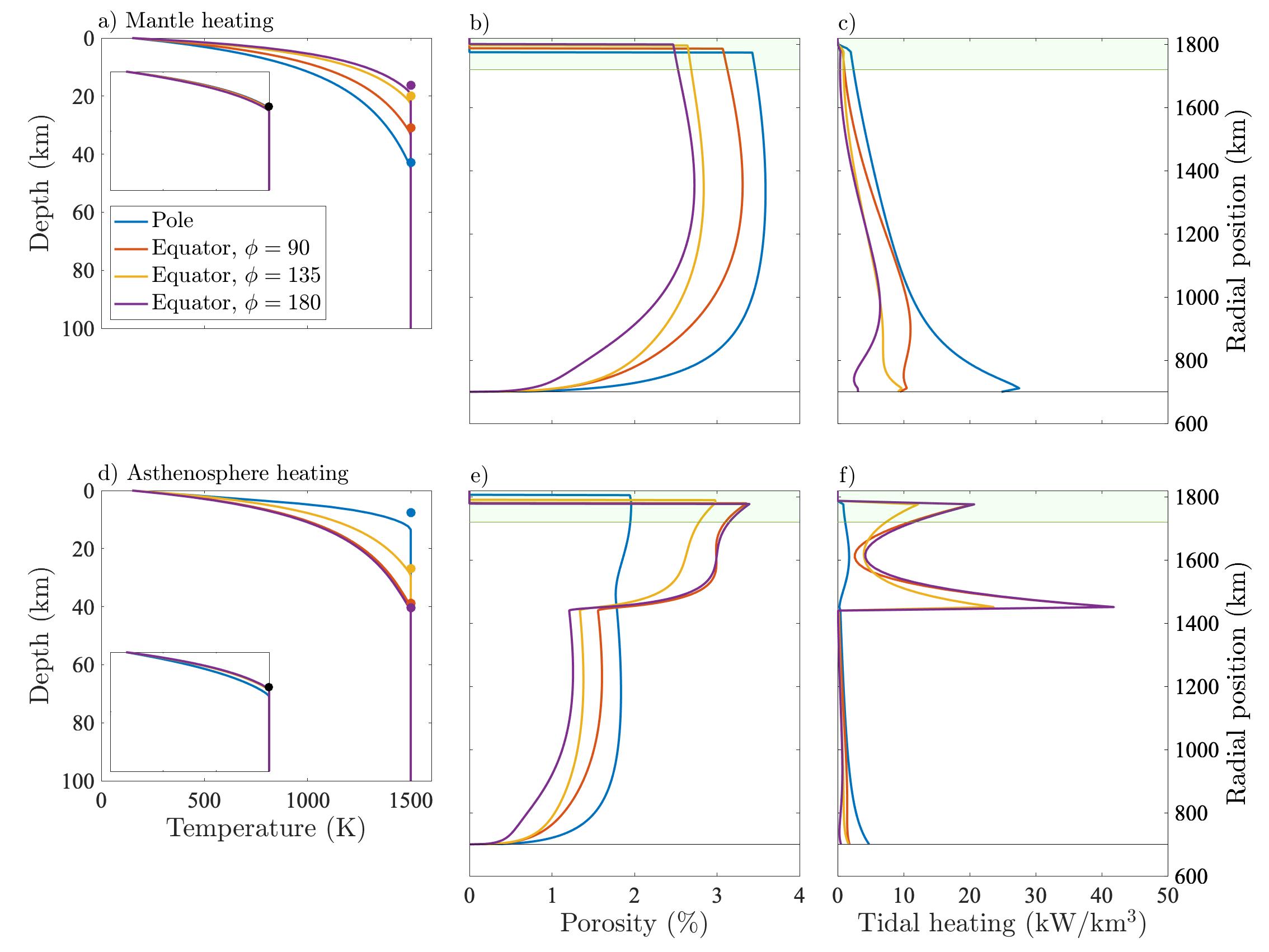}
    \caption{Temperature profiles, mantle porosities, and tidal heating distributions at the poles and three points around the equator, for the mantle heating (a--c) and asthenosphere heating (d--f) models, with constant emplacement rate, $\lambda_{c}=1.66~$Myr$^{-1}$. Panels a and d show the temperature profile in top 100~km layer of Io; this region is indicated by the green strip in the other panels, which show porosity and heating in the whole mantle and lithosphere. Where radially integrated heating rate is highest, melt production and porosity is highest. This results in an increased eruptive flux and the growth of a thicker lithosphere. The insets in panels a and d show the case when emplacement rate is proportional to plumbing system flux, with $\lambda_{q}=0.05~$km$^{-1}$. In this case, lithospheric thicknesses vary weakly (porosity and tidal heating profiles in the mantle are almost exactly the same as constant emplacement rate). Dots indicate estimates of lithospheric thickness using equations \eqref{eq:approx_lamM} (panels a and d) and \eqref{eq:rc_lamq} (insets). Differences between the analytical estimates and the model are caused by conduction, which is neglected in the analytical estimates.}
    \label{fig:radial_profiles}
\end{figure*}

Figure \ref{fig:radial_profiles} shows model solutions for the lithospheric temperature distribution, mantle porosity, and tidal heating distribution at Io's north pole and three points around the equator, for the (coupled) mantle heating model and the (de-coupled) asthenosphere heating model.  In the mantle-heating case (figure \ref{fig:radial_profiles}a--c), heating rate is highest at the poles, and lowest at the sub- and anti-Jovian points, whereas in the asthenosphere-heating case (figure \ref{fig:radial_profiles}d--f), heating rate is highest at the sub- and anti-Jovian points, and lowest at the poles. A higher heating rate leads to increased melt production, though for the permeabilities used here, melt fractions only vary by $\sim 1\%$. Lower permeabilities lead to higher porosities and greater porosity variation between localities \citep{moore_thermal_2001,bierson_test_2016}. Throughout this work, eruption rate and surface heat flux are a proxy for radially integrated heating rate.

The rate of emplacement is controlled by $\lambda_{c}$ and $\lambda_{q}$ (equation \eqref{eq:M}). The values of $\lambda_{c}$ and $\lambda_{q}$ used in this work were chosen to yield an average lithospheric thickness of $\sim 35~$km. Increasing these parameters results in a reduction of the average lithospheric thickness, whilst decreasing them increases the average thickness. This reflects the role of the emplacement rate $M$ in controlling lithospheric thickness, as discussed in \cite{spencer_magmatic_2020}.

An analysis of the equations can be used to obtain a useful analytical approximation for the lithosphere thickness. When the emplacement rate is a constant ($\lambda_{c} \neq 0$ and $\lambda_{q} = 0$), integration of equation \eqref{eq:1Dmass_plum} in the lithosphere yields
\begin{equation}
    q_{p} = \frac{q_{e}R^{2}}{r^{2}}+\frac{\lambda_{c}}{3}\left(\frac{R^{3}}{r^{2}}-r\right), \quad R-l\leq r \leq R,
    \label{eq:qp_analytical}
\end{equation}
where $q_{e} = q_{p}(r=R)$ is the eruption rate. Assuming negligible surface conduction, the eruption rate must be given by a column-wise energy balance as \citep{spencer_magmatic_2020}
\begin{equation}
    q_{e} = \frac{1}{R^{2}(\rho L + \rho C(T_{m}-T_{s}))} \int_{r_{m}}^{R}\psi r^{2}~ \textrm{d}r,
    \label{eq:col_E_bal}
\end{equation}
where the integral is the total tidal heating delivered to the column. From equation \eqref{eq:q_mantle}, the plumbing flux at the base of the lithosphere is
\begin{equation}
	q_{p}(r = R-l) = \frac{1}{(R-l)^{2}\rho L}\int_{r_{m}}^{R-l}\psi r^{2} \textrm{d}r.
    \label{eq:ql_analytical}
\end{equation}
Since negligible tidal heating takes place in the lithosphere (figure \ref{fig:radial_profiles}, note that the green shaded region denotes the upper 100~km, which includes part of the upper mantle where dissipation is not negligible), the integrals in \eqref{eq:col_E_bal} and \eqref{eq:ql_analytical} are essentially identical. Thus, equating \eqref{eq:ql_analytical} with \eqref{eq:qp_analytical} at the base of the lithosphere yields an analytical expression for the lithospheric thickness in terms of the local eruption rate
\begin{equation}
    l = R - R\left(1 - \frac{3q_{e}}{R\lambda_{c}} \frac{C(T_{m}-T_{s})}{L}\right)^{1/3}.
    \label{eq:rc_lamM}
\end{equation}
A Taylor expansion of the term in brackets provides some intuition into this expression. Expanding to the first non-zero term yields
\begin{equation}
l \approx \frac{C(T_{m}-T_{s})}{L}\frac{q_{e}}{\lambda_{c}}.
\label{eq:approx_lamM}
\end{equation}
The thickness of the lithosphere is controlled by the balance between latent heat release in the lithosphere and sensible heat loss at the surface. The greater the temperature difference between erupting lava and the surface, the more heat that must be provided to downwelling material to raise it to its melting point. As the eruption rate increases, material downwells more quickly, and with no corresponding increase in emplacement rate, the thickness of the lithosphere grows. This effect can be seen in the main panels of figure \ref{fig:radial_profiles}. A higher rate of emplacement means that downwelling material is heated more rapidly, reducing the lithospheric thickness. We note that an average lithospheric thickness can be estimated using the modelled global average eruption rate of \cite{spencer_magmatic_2020}.

The insets in panels a and d of figure \ref{fig:radial_profiles} show the lithospheric temperature profiles when emplacement rate is proportional to the plumbing system flux ($\lambda_{c}=0$ and $\lambda_{q}\neq 0$). In this case equation \eqref{eq:1Dmass_plum} can be integrated to give
\begin{equation}
    q_{p} = \frac{R^{2}q_{e}}{r^{2}}\textrm{e}^{\lambda_{q}(R-r)}.
    \label{eq:qp_lamq}
\end{equation}
Again assuming negligible surface conduction and equating equation \eqref{eq:qp_lamq} to the total melt production in the interior (equation \eqref{eq:ql_analytical}) gives an expression for the lithospheric thickness,
\begin{equation}
    l = \frac{1}{\lambda_{q}}\textrm{ln}\left(1 + \frac{C(T_{m}-T_{s})}{L}\right).
    \label{eq:rc_lamq}
\end{equation}
Interestingly, this is independent of the melting rate, so lithospheric thickness is expected to be virtually constant when emplacement rate is proportional to the plumbing system flux. A Taylor expansion of \eqref{eq:rc_lamq} to first order yields equation \eqref{eq:approx_lamM}, but with $q_{e}/\lambda_{c}$ replaced by $1/\lambda_{q}$. The small variations in lithospheric thickness seen in the insets in panels a and d of figure \ref{fig:radial_profiles} are due to conduction (which is neglected in arriving at the estimate, equation \eqref{eq:rc_lamq}), with higher heating rates producing a thinner lithosphere.

\begin{figure*}[ht!]
    \centering
    \includegraphics[width=\linewidth]{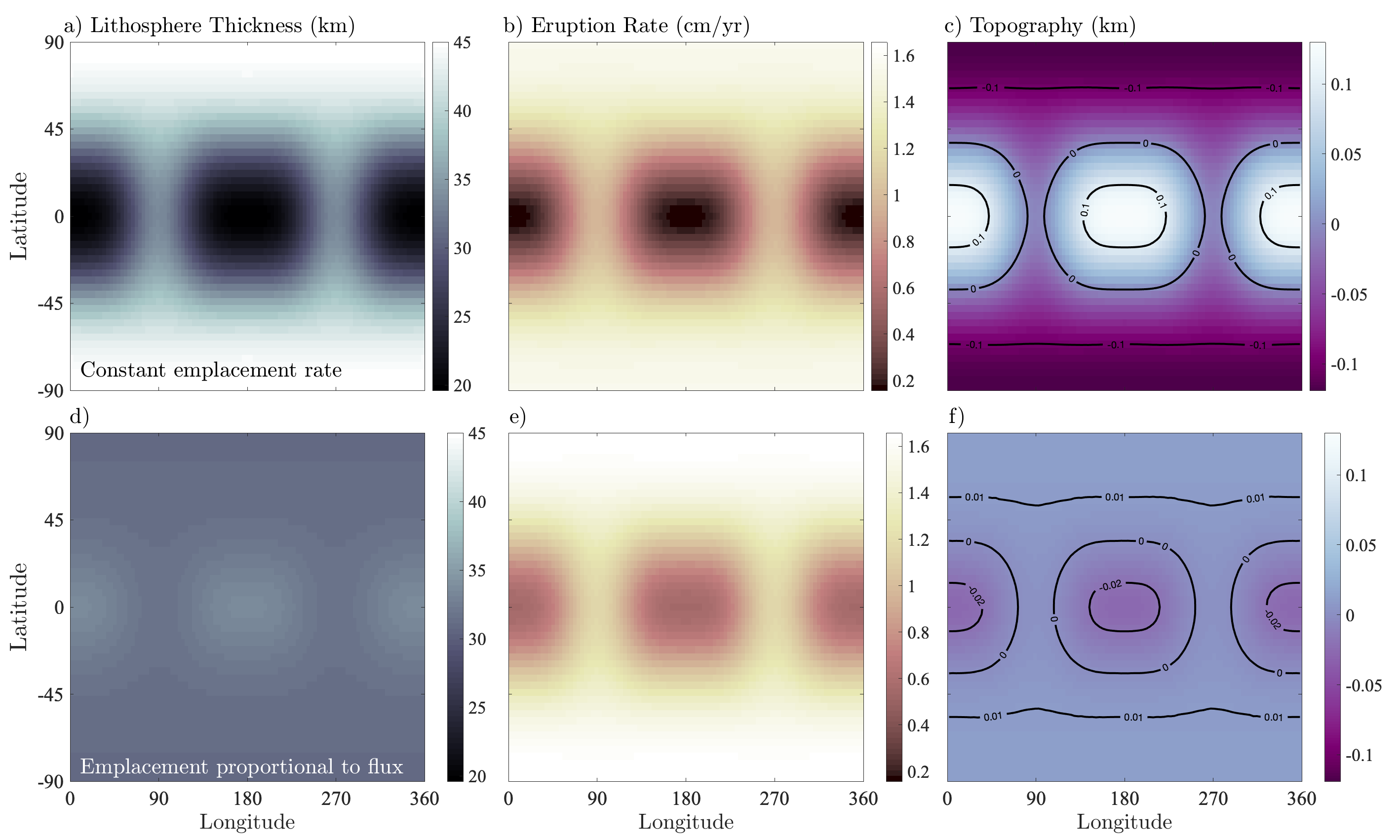}
    \caption{Solutions for lithospheric thickness, eruption rate, and topography in the case of coupled dynamics and tidal heating, assuming the mantle and lithosphere have the same density at equal temperature. Tidal heating is concentrated in the lower mantle in the coupled model, producing maximum eruption rates at the poles (see figure \ref{fig:radial_profiles}). Panels a--c show the case where emplacement rate is constant, and panels d--f show the case where emplacement rate is proportional to the plumbing system flux. Constant emplacement rate predicts a correlation of lithospheric thickness with eruption rate (or heat loss), and topographic lows where heat flux is high. An emplacement rate proportional to plumbing system flux predicts a relatively uniform lithospheric thickness and little long-wavelength topography.}
    \label{fig:mant_heating}
\end{figure*}

Figure \ref{fig:mant_heating} shows lithospheric thickness, eruption rate, and topography as a function of latitude and longitude in the coupled mantle-heating model. The top row of figure \ref{fig:mant_heating} shows the case where emplacement rate is a constant and the bottom row shows the case where emplacement rate is proportional to the plumbing system flux. A constant emplacement rate means that lithospheric thickness correlates with the eruption rate, as specified by equation \eqref{eq:rc_lamM}. Lithospheric thickness varies by about 25~km, with the most pronounced variation being between the thick polar lithosphere and the thin equatorial lithosphere. In figure \ref{fig:mant_heating} we assume that there is no compositionally derived density change at the base of the lithosphere, and so take $\rho_{l} =\rho_{m}$. The lack of a compositional density step means that the cold lithosphere is more dense than the underlying, partially molten mantle; this results in topographic highs where the lithosphere is thinnest. The coupled mantle-heating model with constant emplacement rate predicts long-wavelength topography with an amplitude of about 250~m. In the case where emplacement rate is proportional to the amount of material in the plumbing system, shown in the bottom row of figure \ref{fig:mant_heating}, the lithospheric thickness only varies by a couple of kilometres and the amplitude of long-wavelength topography is $<40~$m. This can be understood through equation \eqref{eq:rc_lamq}; increased heating and the resultant increased eruption rate is balanced by increased emplacement, resulting in an almost uniform lithospheric thickness. In this case, the long-wavelength lithospheric thickness and topography variations are a result of different conductive heat fluxes and so lithospheric thickness is anti-correlated with eruption rate \citep{ross_internal_1990,steinke_tidally_2020}. We stress that the lithospheric-thickness solutions are independent of the topography estimates. The topography estimates rely on an assumption of the compositionally derived density difference (or lack thereof) between the lithosphere and mantle, but the lithospheric thickness estimates do not. An exploration of the effect of varying the mantle density is presented in Appendix A.

\begin{figure*}[ht!]
    \centering
    \includegraphics[width=\linewidth]{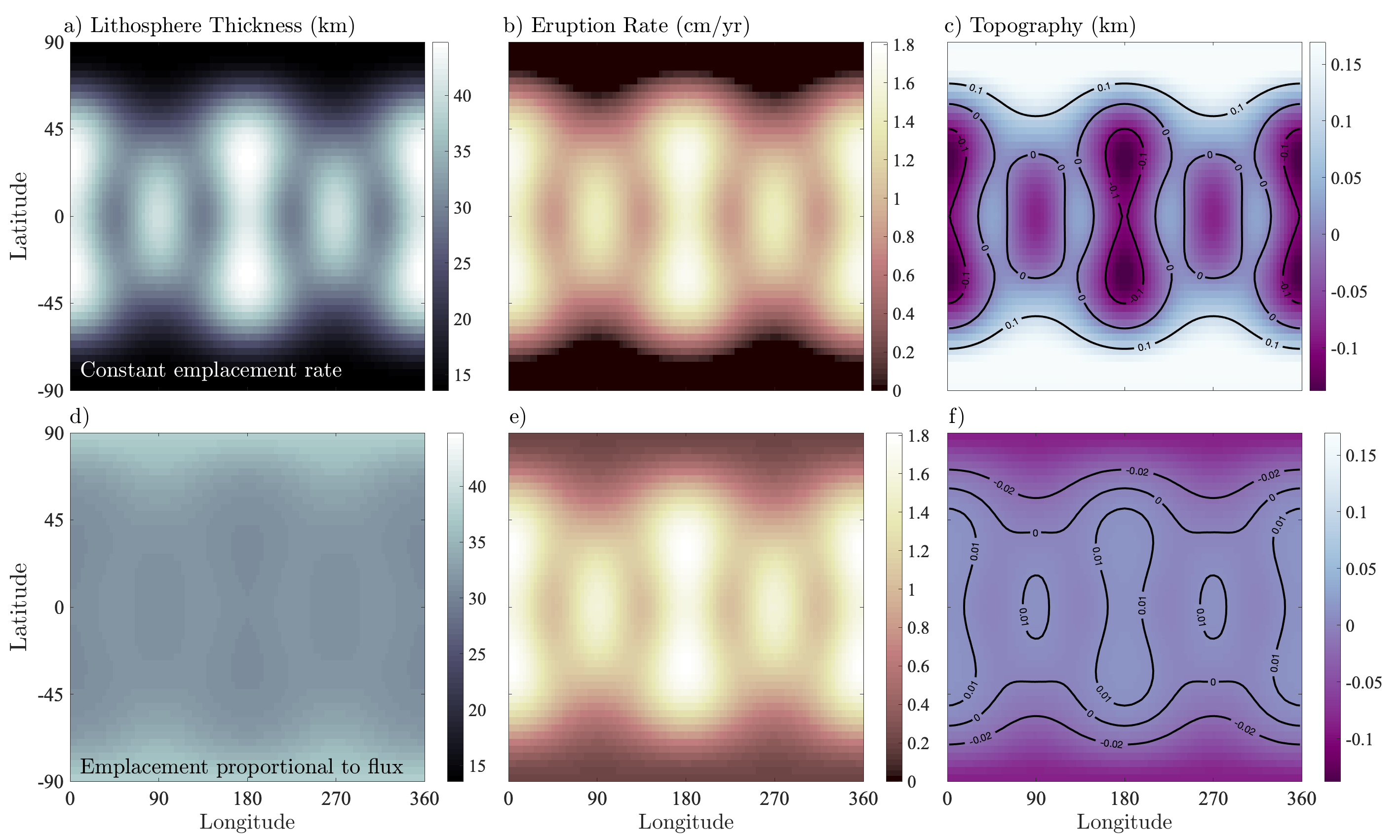}
    \caption{Solutions for lithospheric thickness, eruption rate, and topography in the case of asthenosphere heating. Panels a--c show the case where emplacement rate is constant, and panels d--f show the case where emplacement rate is proportional to the plumbing system flux. Relationships between lithospheric thickness, eruption rate, and topography are the same as in figure \ref{fig:mant_heating}, but patterns and amplitudes are different due to the different heating mode.}
    \label{fig:asth_heating}
\end{figure*}

Figure \ref{fig:asth_heating} shows the same plots as figure \ref{fig:mant_heating}, but for the case of asthenospheric heating. All of the relationships between heating rate, eruption rate, lithospheric thickness, and topography are the same in this case, but the pattern of dissipation and so the pattern of the plotted solutions is different. Asthenospheric heating predicts higher eruption rates at the equator. If emplacement rate is constant, this predicts a thicker lithosphere at the equator (amplitude $\sim 30~$km), and assuming $\rho_{l}=\rho_{m}$, topographic highs at the poles (amplitude $\sim 300~$m). If emplacement rate is proportional to the amount of material in the plumbing system, lithospheric thickness is much more uniform (amplitude $\sim 6~$km) and topography is reduced, with lithospheric thickness variations being controlled by variation in conductive heat fluxes.

Assuming dominantly vertical flow --- a significant assumption that we discuss below --- the global pattern of heat flow should be reflective of the tidal heating distribution, as has been noted elsewhere (e.g., \cite{segatz_tidal_1988,tackley_convection_2001,veeder_io:_2012,davies_map_2015}). The primary means to distinguish between lower mantle and asthenospheric heating models is on the basis of heat flux. Lower mantle heating predicts higher polar heat flux, whereas asthenosphere heating predicts higher equatorial heat flux. With the present dearth of polar observations, this distinction is difficult to make. Rigorous observation of Io's poles is required to understand which mode of heating is more likely to be occurring. However, if the mode of emplacement can be established, lithospheric thickness and topography can serve as a useful proxy for long-timescale heat flux.

\begin{figure*}[ht!]
    \centering
    \includegraphics[width=\linewidth]{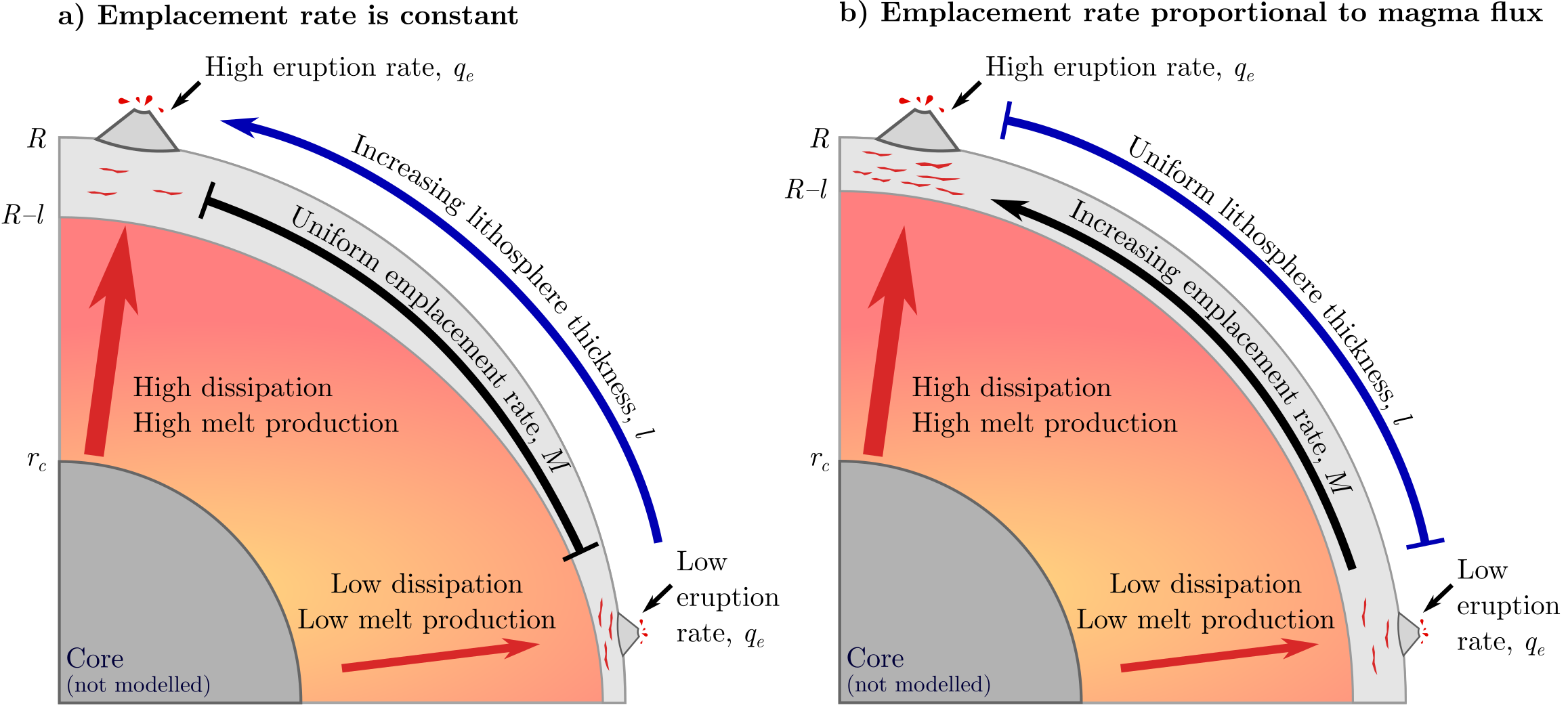}
    \caption{Schematic illustrating the primary results of this work. The lithospheric thickness is correlated with radially integrated heating rate if magmatic intrusions form at a constant rate (panel a), but is approximately uniform (or weakly anti-correlated) if intrusions form at a rate proportional to the flux through volcanic conduits (panel b).}
    \label{fig:tidal_schematic}
\end{figure*}

This work predicts that the long-wavelength variations in lithospheric thickness should either correlate with the long-timescale eruption rate/heat flux, or be weakly anti-correlated, as summarised schematically in figure \ref{fig:tidal_schematic}. In the constant-emplacement-rate model, we predict that lithospheric thickness correlates with eruption rate. An explanation for why emplacement would be independent of magma flux is that volcanic conduits are not formed by magma pressure at depth, but rather tectonic processes in the lithosphere. Io's eruption-and-burial tectonics are thought to form mountains by thrust faulting \citep{mckinnon_chaos_2001,kirchoff_formation_2009}. If, for example, such faults can act as conduits for magma ascent, freezing of ascending magma on their walls may be largely independent of the flux through the conduit. Alternatively, in the flux-proportional emplacement rate model, we predict that long-wavelength lithospheric thickness varies by only a few kilometers, and is weakly anti-correlated with heat flux. A rationale for why emplacement rate would be proportional to volcanic plumbing flux may be that volcanic conduits are created by overpressured magma at the base of the lithosphere. It is plausible that higher melt production in the interior would lead to a larger number of conduits. If magma in each of these conduits has a chance of stalling within the lithosphere, this would imply a positive relationship between lithospheric magma flux and emplacement rate.

The flux-proportional emplacement rate model makes predictions for variations in lithospheric thickness that are similar to the results of \cite{steinke_tidally_2020}. When comparing this work to \cite{steinke_tidally_2020}, it is important to note that whilst both can predict a conductive control on lithospheric thickness variations, the controls on the absolute values of lithospheric thickness are different. In this work the lithospheric thickness is primarily controlled by the rate of magmatic emplacement, whereas in \cite{steinke_tidally_2020} the lithospheric thickness is controlled entirely by conduction through a stagnant lid. To address the relative importance of convective heat transport in the mantle likely requires a model that couples two-phase flow and convection, a significant challenge due to the different timescales on which these processes operate.

The proposed link between lithospheric thickness and topography provides a means of relating more readily-obtainable observations (topography) to the predictions of lithospheric thickness in works like this one and \cite{steinke_tidally_2020}. However, to quantify this prediction requires an additional assumption about the relative densities of the lithosphere and mantle. \cite{spencer_compositional_2020} demonstrated that in Io's top-down tectonics, the lithosphere is expected to be efficiently recycled into the mantle, resulting in the lithosphere and upper mantle having the same composition. This is the case presented in figures \ref{fig:mant_heating} -- \ref{fig:asth_heating}, where the densities of the mantle and lithosphere are the same when their temperatures are equal. It is nonetheless plausible that as erupta is buried through the lithosphere, fusible components are melted and mobilised first, which may in turn lead to a density stratification. The extent of this effect is perhaps small, however, given the expected mafic nature of Io's lithosphere. In Appendix A we show that if Io does have a distinct layer of different chemical density, its effect on topography depends on the density difference $\rho_{m}-\rho_{l}$. At a critical density difference of $\sim 30~$kg/m$^{3}$, the compositional density effect cancels out the temperature dependence of density, resulting in no topography. If $\rho_{m}-\rho_{l}$ is sufficiently large, topography is inverted from that presented in figures \ref{fig:mant_heating} -- \ref{fig:asth_heating}. If the density difference can indeed be estimated, topography observations can be compared to eruption rates and volcanic heat fluxes to clarify the heat-transfer and emplacement mechanisms in the lithosphere. Alongside recent work that demonstrates a way to constrain interior structure from libration amplitudes \citep{vanhoolst_librations_2020}, this provides a means to investigate Io's interior structure and heating distribution. 

Our isostasy calculations assume that compositional variation within the lithosphere can be approximated by a density step at the base of the lithosphere. It is likely that the compositional profile in the lithosphere is complex, reflecting shallow magma fractionation, sulfur cycling, and other processes. If the vertical structure of the lithosphere is approximately uniform with latitude and longitude, and simply scaled to lithospheric thickness, the results of this work should be largely unchanged. If, however, there is significant variation in lithosphere composition with latitude and longitude, the applicability of the isostatic model presented here would be reduced. It is not clear, however, that any such variation would mirror the degree-two tidal forcing, and so may average out on the long wavelengths considered here.

\cite{white_new_2014} created a partial stereo-topographic DEM of Io that found a system of longitudinally arranged alternating basins and swells near the equator, with amplitudes $\sim$ 1--2~km and a wavelength $\sim 400~$km. This large amplitude may imply that compositional density differences are important in controlling topography (see Appendix \ref{Appendix}), or that dynamic topography caused by upwelling mantle plumes is significant \citep{tackley_three-dimensional_2001}. It is important to note however that there are considerable discrepancies between stereo-derived and limb-profile-derived long-wavelength topography \citep{white_new_2014}, and hence that long-wavelength topography is not well constrained. Further, the long-wavelength, isostatic topography described here may be difficult to disentangle from tidal and rotational deformation. Efforts are generally made to remove tidal and rotational effects from global topographic maps, but this process may also inadvertently remove all or part of the topography described here. Improved observations of long-wavelength topography, particularly in the polar regions, as well as a means of disentangling different contributions to long-wavelength topography are required to make robust comparisons between modelled topography and data.

A primary limitation of this work is the neglect of lateral flow in either the lithosphere or mantle. Differences in lithospheric thickness are expected to be counteracted by deformation of the lithosphere. Such calculations are common in studies of the ice shells of icy satellites \citep{stevenson_limits_2000,nimmo_estimates_2001,nimmo_non-newtonian_2004}, where there is a clear rheological and density transition at the base of the shell. The application of such a model to Io is not straightforward because rheological and density transitions are expected to be more gradual \citep{spencer_compositional_2020}. It is not clear whether there is an easily defined petrological `crust' of Io. Nonetheless such lateral flow is possible, and would be best investigated by a two-dimensional model of upper Io. Lateral flow is also possible in the partially molten mantle. Pressure gradients would be expected to drive flow of the mobile magma phase. Pressure gradients could be produced by processes such as different melting rates or spatially variable extraction rates to the lithosphere. An investigation of lateral melt flow would likely require a two-dimensional model of the partially molten mantle. Here we simply note that the relationships proposed in this model are expected to hold if vertical motion is much greater than lateral motion, as generally expected in Io's eruption-and-burial tectonics at long wavelengths.

\section{Conclusions}
We have demonstrated how spatially variable tidal heating leads to long-wavelength variations in lithospheric thickness in models of magmatic segregation and volcanic eruptions. Our models predict that such variations are controlled by how magma intrudes into the lithosphere. If permanent magmatic intrusions form at a rate independent of the magma flux through volcanic conduits, the lithosphere should be thickest where tidal heating is greatest. In this case the lithopshere thickness can vary by 10s of km. If, however, magmatic intrusions form at a rate proportional to the magma flux through volcanic conduits, lithospheric thickness will only vary by a few km, and will be anti-correlated with eruption rates. We also predict that if density differences are predominantly derived from temperature differences, then areas of thin lithosphere will sit on topographic highs. Improved observational constraints on eruption rates, heat fluxes, and long-wavelength topography, particularly at Io's poles, will help distinguish between different models for the controls on lithospheric thickness.

\appendix
\section{Appendix A --- Topographic effects of a petrologically distinct crust}
\label{Appendix}
The isostasy calculations presented in this work require an assumption about the chemical densities of the lithosphere and the underlying mantle. If the lithosphere and upper mantle have different chemical densities, then this will affect topography. In Io's top-down, heat-pipe tectonics, compositions are expected to become increasingly refractory with depth. \cite{keszthelyi_magmatic_1997} proposed that the near surface would be fusible and composed of low density, silica rich components. This view fell out of favour when improved observations illuminated the mafic to ultra-mafic nature of the lithosphere \citep{keszthelyi_new_2007}. \cite{spencer_compositional_2020} demonstrated that efficient recycling of erupted lavas back into the mantle prevents the mafic near-surface from significantly differentiating; they proposed that the lithosphere and upper mantle have approximately the same composition. Heat-pipe tectonics thus appears to result in a relatively uniform composition in the near surface; Io may well lack a petrologically distinct crust.

Figures \ref{fig:mant_heating} -- \ref{fig:asth_heating} incorporate this assumption, taking the lithosphere and mantle to have the same density at equal temperatures ($\rho_{m}=\rho_{l}$). It is plausible, however, that a degree of compositional differentiation does take place in the near surface, with more fusible material being mobilised first, which may in turn result in a density difference ($\rho_{m}\neq \rho_{l}$). This process likely doesn't produce large density differences for the reasons described above, but may still play a significant role in controlling topography.

\begin{figure*}
\centering
\includegraphics[scale=0.5]{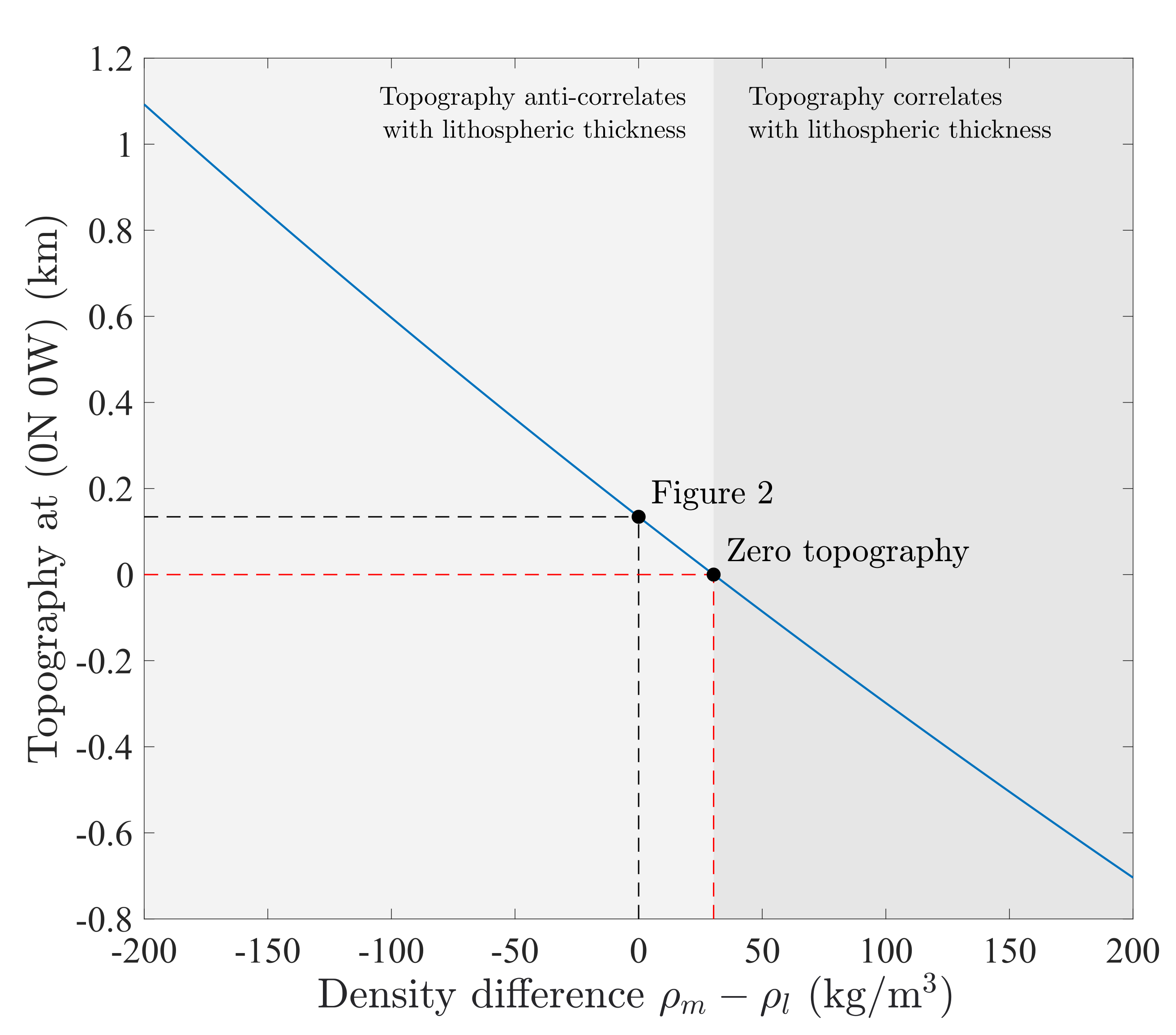}
\caption{Topography at the sub- or anti-Jovian point ($0$N $0/180$W) for the coupled, mantle heating model as a function of the chemical density difference $\rho_{m}-\rho_{l}$ between the upper mantle and lithosphere. If the lithosphere is buoyant with respect to the mantle, topography is expected to correlate with lithospheric thickness. For a density difference of $\sim 30~$kg/m$^{3}$ topography derived from the low lithospheric temperature is cancelled out, resulting in no topography. Figures \ref{fig:mant_heating} -- \ref{fig:asth_heating} assumed $\rho_{m}=\rho_{l}$.}
\label{fig:rhom_vary}
\end{figure*}

Figure \ref{fig:rhom_vary} shows the topography at the sub- or anti-Jovian point for the coupled, mantle-heating model as a function of the density difference $\rho_{m}-\rho_{l}$. Figure \ref{fig:rhom_vary} shows that if the mantle is $\sim 30~$kg/m$^{3}$ more dense than the lithosphere, then the temperature dependence of density in the lithosphere is cancelled out, resulting in no topography. If the density difference is greater than this then the topography patterns in figures \ref{fig:mant_heating} -- \ref{fig:asth_heating} are inverted.

\printcredits

%% Loading bibliography style file
%\bibliographystyle{model1-num-names}
\bibliographystyle{cas-model2-names}

% Loading bibliography database
\bibliography{Oxford,Software}

\end{document}